\documentclass[prl,twocolumn]{revtex4}
\usepackage{graphicx}
\usepackage{amssymb,amsmath}
\usepackage{enumerate}

\newcommand{\be}{\begin{eqnarray}}
\newcommand{\ee}{\end{eqnarray}}

\begin{document}

\title{Chaotic and turbulent mixing of passive scalar}

\author{A. Bershadskii}

\affiliation{
ICAR, P.O. Box 31155, Jerusalem 91000, Israel
}

\begin{abstract}
Spatio-temporal deterministic chaos at small Taylor-Reynolds numbers $Re_{\lambda} \lesssim 40$ and distributed chaos at turbulent $Re_{\lambda} \gtrsim 40$ in passive scalar dynamics have been studied using results of direct numerical simulations of homogeneous incompressible flows (with and without mean gradient of the passive scalar) for $8 \leq Re_{\lambda} < 700$ and of a reacting turbulent mixing layer. It is shown that the deterministic chaos in the passive scalar fluctuations at the small $Re_{\lambda}$ is characterized by exponential spatial (wavenumber) spectrum: $E(k) \propto \exp-(k/k_c)$, whereas the distributed chaos at turbulent $Re_{\lambda}$ is characterized by stretched exponential spectrum $E(k) \propto \exp-(k/k_{\beta})^{3/4}$. The Birkhoff-Saffman invariant related to the momentum conservation and, due to the Noether theorem, to the spatial homogeneity has been used as a theoretical basis for this stretched exponential spectrum.  Although the $k_c$ and $k_{\beta}$ represent the large-scale structures a relevance of the Batchelor scale $k_{bat}$ has been established as well: the normalized values $k_c/k_{bat}$ and $k_{\beta}/k_{bat}$ exhibit universality.

\end{abstract}

\maketitle

\section{Deterministic spatio-temporal chaos}
\begin{figure}
\begin{center}
\includegraphics[width=8cm \vspace{-0.3cm}]{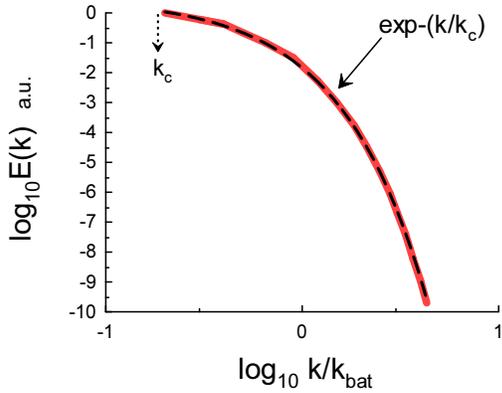}\vspace{-4cm}
\caption{\label{fig1} Power spectrum of passive scalar fluctuations for isotropic homogeneous (steady) spatio-temporal fluid chaos ($Re_{\lambda} = 8$ and $Sc = 1$).} 
\end{center}
\end{figure}
\begin{figure}
\begin{center}
\includegraphics[width=8cm \vspace{-0.3cm}]{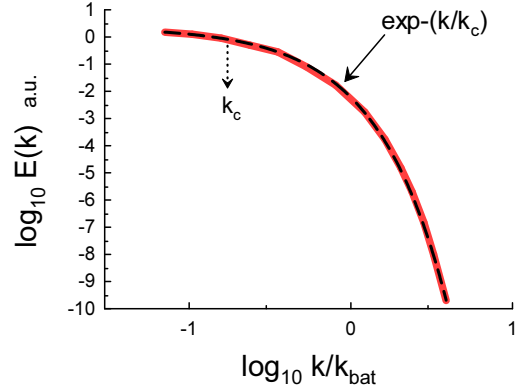}\vspace{-3.9cm}
\caption{As in the Fig. 1 but for $Sc = 8$.}
\end{center}
\end{figure}

 Smooth dynamical systems with temporal chaos and compact strange attractors  have as a rule exponential {\it frequency} spectra \cite{oh}-\cite{mm}. For Hamiltonian systems the smoothness results in the stretched exponential frequency spectra, whereas violation of the smoothness results in power-law (scaling) frequency spectra \cite{b2}.  One can expect that for the systems described by equations with partial derivatives and spatio-temporal deterministic chaos the smoothness should result in the spatial (wavenumber) exponential spectra as well. \\
 
 The evolution equation for a passive scalar $\theta$
$$
\partial_t \theta + \bf{v}\cdot\nabla\theta= \kappa\nabla^2\theta+f_{\theta}  \eqno{(1)}
$$
with velocity field given by the incompressible Navier-Stokes equations
$$  
  \partial_t {\bf v} + {\bf v}\cdot \nabla {\bf v} = -\nabla p +\nu \nabla^2 {\bf v}+\mathbf{f}_{\bf v}, \eqno{(2)}
$$
$$  
  \nabla \cdot \bf{v}=0   \eqno{(3)}
$$  
can provide a good example for small Reynolds numbers, when one can expect appearance of the deterministic spatio-temporal chaos. From theoretical point of view and with a perspective to continue the study on the turbulent flows (see below) the isotropic and homogeneous case is the most suitable one. 

  In order to attain a statistically stationary (steady) state for velocity and passive scalar fields different forcing methods are used in the direct numerical simulations. For the velocity field they are usually spectral or linear \cite{lun}-\cite{dy}, whereas for the passive scalar field a uniform mean gradient or a random field at low wavenumbers are usually applied \cite{bdy}-\cite{yds}. The periodic boundary conditions are also the most common ones in the direct numerical simulations. \\

  Figures 1 and 2 show the spatial (wavenumber) power spectra for such passive scalar mixing for the Taylor-Reynolds number $Re_{\lambda}=8$ and for the Schmidt number $Sc=1$ and $8$ respectively. The spectral data were taken from Fig. 1a of the Ref. \cite{dsy} where results of a direct numerical simulation of the passive scalar mixing in the isotropic homogeneous fluid motion were reported (for the velocity field two related spectral forcing methods and for the passive scalar field the mean gradient forcing method were used). The $k_{bat}$ is the Batchelor wavenumber. The dashed curves indicate the exponential power spectrum $E(k)$ of the passive scalar $\theta$ 
$$
 E(k) \propto \exp-(k/k_c) \eqno{(4)}
$$ 
$k$ is the wavenumber. The dotted arrows indicate position of the characteristic scale $k_c$ for each case: $k_c \simeq 0.18~k_{bat}$ for $Sc =1$ and $ k_c \simeq 0.17 ~k_{bat}$ for $Sc=8$. \\

  Since the $Re_{\lambda} =8$ is small (see next section) one can expect that in this case a spatio-temporal deterministic chaos takes place for the both values of the Schmidt number. 

\section{Distributed chaos and homogeneous turbulence}

   With $R_{\lambda} $ increasing beyond $Re_{\lambda} \simeq 40$ the isotropic homogeneous fluid motion becomes turbulent (see, for instance, Refs. \cite{s},\cite{sb} and references therein). At this transition the parameter $k_c$ fluctuates strongly and one needs in an ensemble average 
$$
E(k) \propto \int_0^{\infty} P(k_c) \exp -(k/k_c)~ dk_c  \eqno{(5)}
$$
in order to obtain the spatial spectra. Here $P(k_c)$ is a corresponding probability distribution of the $k_c$. Since the system is still smooth it is reasonable to seek the spectrum in a stretched exponential form
$$
E(k) \propto \exp-(k/k_{\beta})^{\beta},   \eqno{(6)}
$$ 
the $k_{\beta}$ is a constant (cf. Refs. \cite{b2},\cite{b3}). Now the task is to find value of the parameter $\beta$, if there is a universal one related to the fundamental properties of the system. \\

  For this purpose one can use asymptotic properties of the distribution $P(k_c)$ at $k_c \rightarrow \infty$. On the one hand, it follows from the Eqs. (5) and (6) that the asymptotic of $P(k_c)$ at $k_c \rightarrow \infty$ has the form \cite{jon}
$$
P(k_c) \propto k_c^{-1 + \beta/[2(1-\beta)]}~\exp(-bk_c^{\beta/(1-\beta)}), \eqno{(7)}
$$
$b$ is a constant. On the other hand, asymptotic distribution $P(k_c)$ can be found from the background physics of the system. It is known that the Birkhoff-Saffman integral 
$$
I = \int_0^{\infty} B_{LL} (r) r^2dr  \eqno{(8)}
$$
($B_{LL}$ is longitudinal correlation function of the velocity field) is an invariant of the Navier-Stokes equations  related to the momentum conservation law and, due to the Noether's theorem, to the spatial homogeneity \cite{b3},\cite{d}-\cite{js}.  The Birkhoff-Saffman integral can be estimated as 
$$
I \propto v_c^2~ l_c^3   \eqno{(9)}
$$
where $v_c$ is a characteristic velocity and $l_c$ is a characteristic spatial scale. Using relationship $l_c \propto k_c^{-1}$ one can obtain from the Eq. (9)
$$
k_c \propto I^{-1/3}~v_c^{2/3}   \eqno{(10)}
$$
 If distribution of $v_c$ is a Gaussian one (with zero mean), then the asymptotic distribution of the $k_c$ can be obtained from the Eq. (10)
$$
P(k_c ) \propto k_c^{1/2}~\exp-\left(\frac{k_c}{K}\right)^3  \eqno{(11)}
$$   
where $K$ is a constant. Then comparing Eq. (7) with Eq. (11) one obtains
$$
\beta =3/4  \eqno{(12)}
$$
for the homogeneous distributed chaos (turbulence).

\begin{figure}
\begin{center}
\includegraphics[width=8cm \vspace{-1cm}]{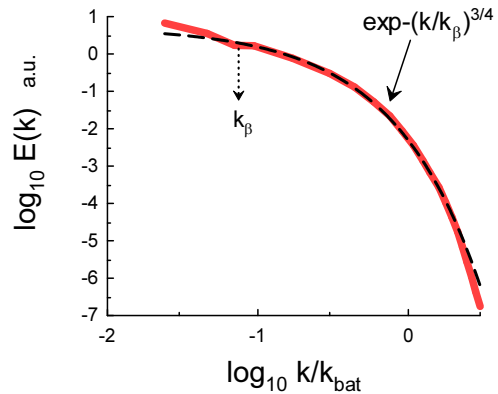}\vspace{-3.7cm}
\caption{ Power spectrum of passive scalar fluctuations for steady homogeneous spatio-temporal distributed chaos with $Re_{\lambda} = 55$ and $Sc = 1$. } 
\end{center}
\end{figure}
\section{Direct numerical simulations of the homogeneous turbulence}

     In Ref. \cite{cvb} the steady homogeneous turbulence was generated by a linear forcing method. Figure 3 shows the spatial (wavenumber) power spectrum for the passive scalar fluctuations for $Re_{\lambda}= 55$ and for the Schmidt number $Sc=1$ (cf. Fig. 1 with the same value of the Schmidt number but with the $Re_{\lambda} = 8$). The spectral data for this figure were taken from Fig. 3a of the Ref. \cite{cvb} (corresponding to the mean gradient forcing of the passive scalar). The dashed curve indicates the stretched exponential spectrum Eq. (6) with the $\beta =3/4$ - Eq. (12). Value of the scale
$$
k_{\beta} \simeq 0.075~ k_{bat}    \eqno{(13)}
$$
This scale was already mentioned in the Ref. \cite{b3} as a possible universal one (see also below).\\

   Now let us return to the direct numerical simulation reported in the Ref. \cite{dsy} but for the turbulent value of the $Re_{\lambda}$. Figure 4 shows the spatial (wavenumber) power spectrum for the passive scalar fluctuations for $Re_{\lambda}= 140$ and for the Schmidt number $Sc=1$ (cf. Fig. 1 with the same value of the Schmidt number but with the $Re_{\lambda} = 8$). The spectral data for this figure were taken from Fig. 1b of the Ref. \cite{dsy} (corresponding to the mean gradient forcing of the passive scalar). The dashed curve indicates the stretched exponential spectrum Eq. (6) with the $\beta =3/4$ - Eq. (12). Again the value of $k_{\beta} \simeq 0.075~k_{bat}$ (cf. Eq. (13)).\\
    
  Figure 5 shows the spatial (wavenumber) power spectrum for the passive scalar fluctuations for $Re_{\lambda}= 140$ and for the Schmidt number $Sc=0.125$. The spectral data for this figure were taken from Fig. 1b of the Ref. \cite{dsy}. The dashed curve indicates the stretched exponential spectrum Eq. (6) with the $\beta =3/4$ and the value of $k_{\beta} \simeq 0.073~k_{bat}$ (cf. Eq. (13)). \\ 
 \begin{figure}
\begin{center}
\includegraphics[width=8cm \vspace{-0.7cm}]{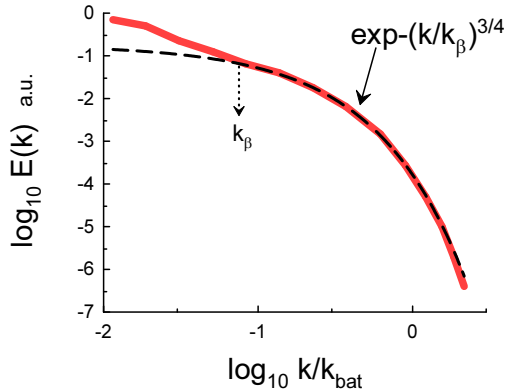}\vspace{-3.9cm}
\caption{ The same as in Fig. 3 but for $Re_{\lambda} = 140$ and $Sc = 1$. } 
\end{center}
\end{figure}
\begin{figure}
\begin{center}
\includegraphics[width=8cm \vspace{-0.58cm}]{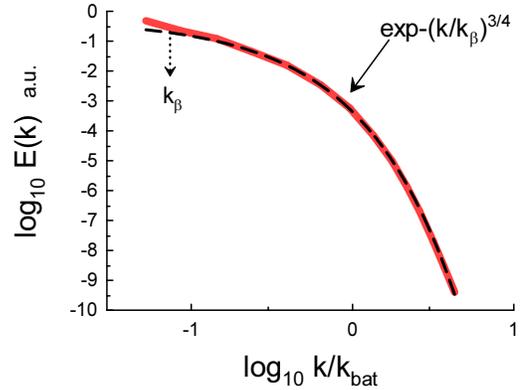}\vspace{-4.15cm}
\caption{\label{fig2} The same as in Fig. 4 but for $Re_{\lambda} = 140$ and $Sc = 0.125$. } 
\end{center}
\end{figure}
\begin{figure}
\begin{center}
\includegraphics[width=8cm \vspace{-0.3cm}]{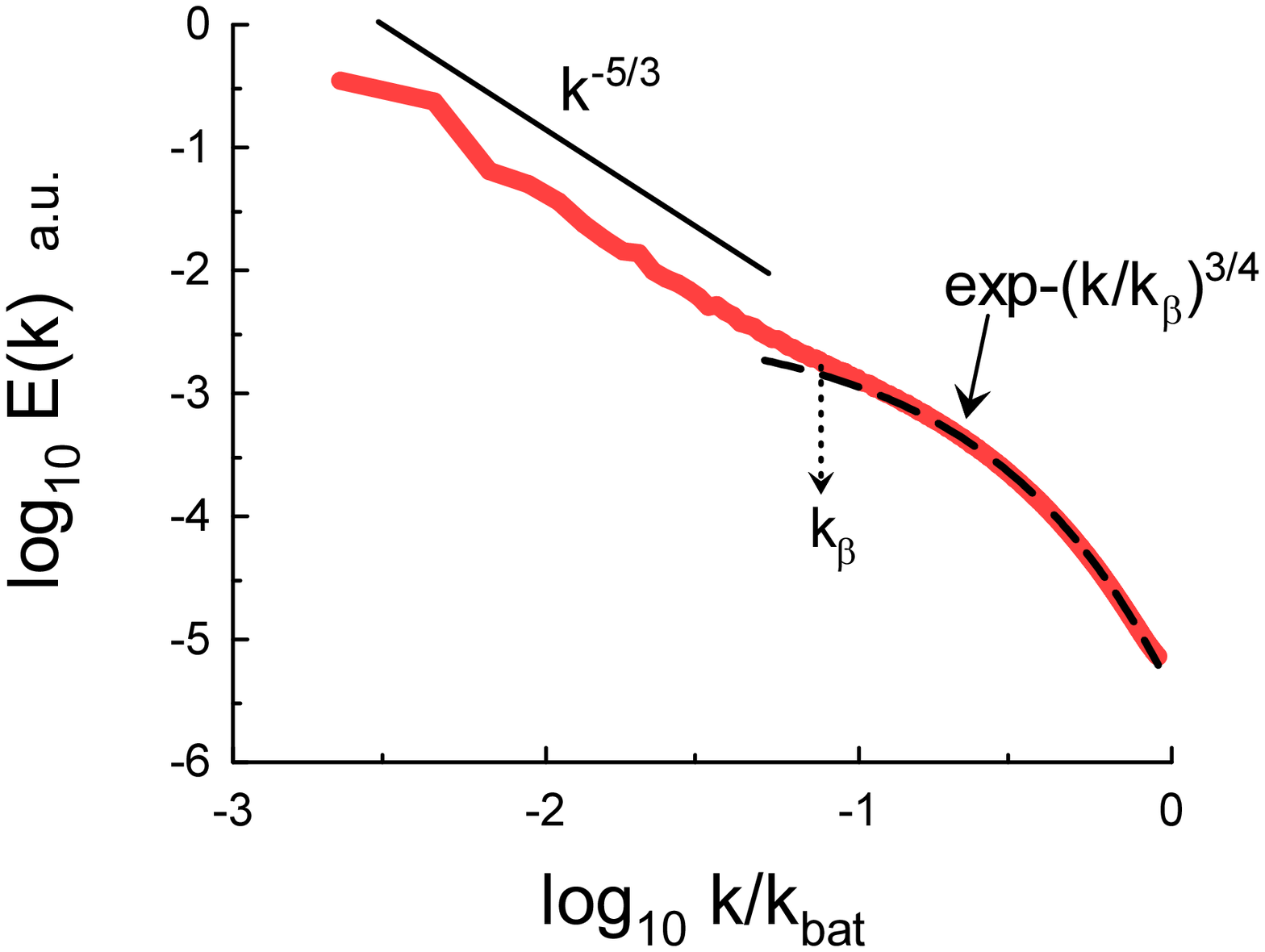}\vspace{-3.55cm}
\caption{\label{fig2} Power spectrum of the passive scalar fluctuations for the $Re_{\lambda}=427$ and the $Sc=1$ } 
\end{center}
\end{figure}  
 \begin{figure}
\begin{center}
\includegraphics[width=8cm \vspace{-0.83cm}]{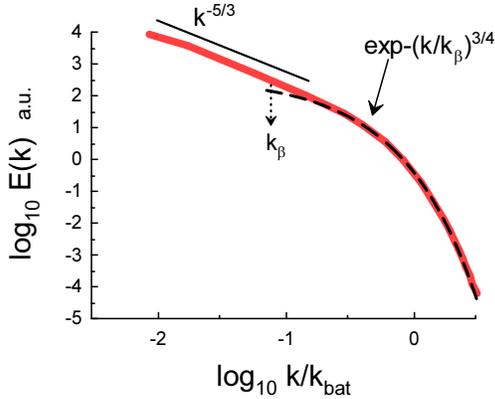}\vspace{-3.95cm}
\caption{\label{fig2} Power spectrum of the passive scalar fluctuations for the $Re_{\lambda}=700$ and the $Sc=0.125$ } 
\end{center}
\end{figure}  

 With a further increase of the $Re_{\lambda}$ the smoothing action of the molecular viscosity becomes insufficient to smooth out the fluid motion at large spatial scales (small wavenumbers) and a power-law spectrum can appear (cf. Ref. \cite{b2}) near the distributed chaos range. Figure 6 shows power spectrum of the passive scalar fluctuations for the $Re_{\lambda}=427$ and the $Sc=1$ observed in a direct numerical simulation of the steady isotropic homogeneous turbulence \cite{wg}. The $f_{\theta}$ (scalar source) and $\mathbf{f}_{\bf v}$ (Gaussian random force) in the Eqs. (1-2) are delta-correlated in time and have been added in the low-wavenumbers range. The straight line with the slope '-5/3' in the log-log scales is drawn for reference of the Obukhov-Corrsin power law \cite{my}. The dashed curve indicates the stretched exponential spectrum Eq. (6) with the $\beta =3/4$ Eq. (12) in the distributed chaos range of scales. And again the value of $k_{\beta} \simeq 0.075~k_{bat}$ (cf. Eq. (13)). \\

   In paper Ref. \cite{yds} results of a direct numerical simulation with $Re_{\lambda} \simeq 700$ were reported. In this DNS the isotropic homogeneous velocity field was forced according to the stochastic scheme suggested in Ref. \cite{ep} whereas the passive scalar field was forced by a uniform mean scalar gradient. As for the previous DNS periodic boundary conditions were  applied. Figure 7 shows power spectrum of the passive scalar fluctuations for the steady isotropic homogeneous turbulence  with $Re_{\lambda} \simeq 700$ and $Sc=0.125$.  The spectral data for this figure were taken from Fig. 2 of the Ref. \cite{yds}. The straight line with the slope '-5/3' in the log-log scales is drawn for reference of the Obukhov-Corrsin power law \cite{my}. The dashed curve indicates the stretched exponential spectrum Eq. (6) with the $\beta =3/4$ Eq. (12) in the distributed chaos range of scales. The value of $k_{\beta} \simeq 0.076~k_{bat}$ (cf. Eq. (13)).\\

  Finally, in the paper Ref. \cite{ss} results of a direct numerical simulation of a mixing (decaying) passive scalar blob in a steady isotropic homogeneous turbulence at $Re_{\lambda} =64$ and $Sc=8$ are reported. Figure 8 shows the spatial (wavenumber) power spectrum for the passive scalar fluctuations at max. computational time $T/\tau_{\eta} =11.9$, where $\tau_{\eta} = \eta^2/\nu$, $\eta$ is Kolmogorov scale, $\nu$ is viscosity. The spectral data for this figure were taken from Fig. 2b of the Ref. \cite{ss}. The dashed curve indicates the stretched exponential spectrum Eq. (6) with the $\beta =3/4$ Eq. (12). \\
\begin{figure}
\begin{center}
\includegraphics[width=8cm \vspace{-0.9cm}]{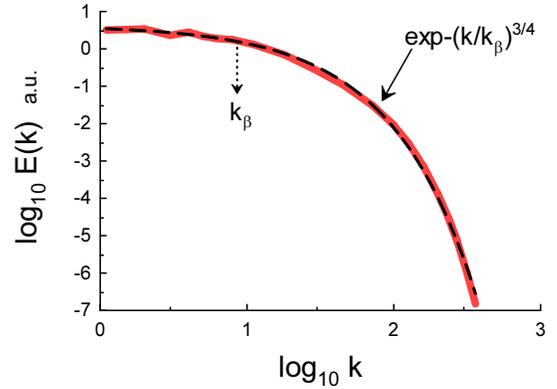}\vspace{-3.85cm}
\caption{  Power spectrum for the passive scalar fluctuations in a decaying passive scalar blob in a steady isotropic homogeneous turbulence at $Re_{\lambda} =64$ and $Sc=8$. } 
\end{center}
\end{figure}    
\begin{figure}
\begin{center}
\includegraphics[width=6.5cm \vspace{-0.1cm}]{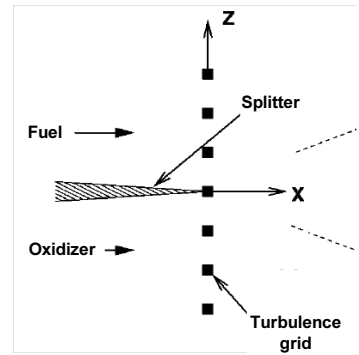}\vspace{-3cm}
\caption{A reacting mixing layer behind the turbulence generating grid. } 
\end{center}
\end{figure}

\section{Turbulent reacting mixing layer}

  An interesting numerical simulation of a passive reacting scalar mixing layer was reported in Ref. \cite{krk}. In this numerical simulation a passive chemical reaction of two  initially separated  species was studied in a grid generated (shear-free) turbulence with $Sc=1$. For the passive chemical reactions the heat released by the reactions 
  has no practical effect on the hydrodynamics.  Dilute oxidant and fuel are injected through two separate halves of a
turbulence grid in a wind tunnel. Then mixing and reaction occur in presumably isotropic and homogeneous decaying turbulence behind the grid: Fig. 9 (adapted from the Ref. \cite{krk}). To ensure that the passive scalar and velocity are accurately simulated an iterative matching with several well known laboratory experiments was provided. The Taylor hypothesis, relating the spatial and temporal statistics \cite{my}, was used for this matching. \\

  Figure 10 shows the spatial (wavenumber) power spectrum for the conserved scalar fluctuations at distance $x/M =231$ behind the grid ($M$ is the spacing of the grid generating turbulence), and at $z=0$. Average of the spectra over the different planes was applied. The spectral data for this figure were taken from Fig. 12 of the Ref. \cite{krk}. The dashed curve indicates the stretched exponential spectrum Eq. (6) with the $\beta =3/4$ Eq. (12). The value of $k_{\beta}$ corresponds to the largest spatial scales (the dotted arrow in the Fig. 10) and, consequently, the entire distributed chaos is tuned to these scales (cf. Fig. 1 with analogous behaviour of the deterministic chaos). \\
\begin{figure}
\begin{center}
\includegraphics[width=8cm \vspace{-1cm}]{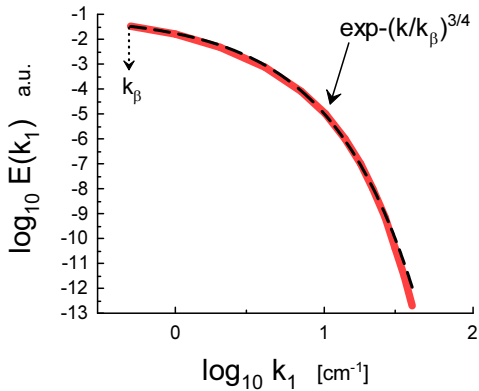}\vspace{-4cm}
\caption{ Power spectrum for the passive (conserved) scalar fluctuations at distance $x/M =231$ behind the grid and at $z=0$ ($Sc= 1$). } 
\end{center}
\end{figure}

\section{Acknowledgement}

I thank  T. Gotoh, and T. Watanabe for sharing 
their data, and A. Pikovsky for stimulating discussion.


\begin{thebibliography}{99}

\bibitem{oh} N. Ohtomo, K. Tokiwano, Y. Tanaka et. al., J. Phys. Soc.
Jpn. {\bf 64} 1104 (1995).
\bibitem{fm}U. Frisch and R. Morf, Phys. Rev., {\bf 23}, 2673 (1981).
\bibitem{f} J.D. Farmer, Physica D, {\bf 4}, 366 (1982).
\bibitem{sw}A. Brandstater and H.L. Swinney, Phys. Rev. A {\bf 35}, 2207 (1987).
\bibitem{sig} D.E. Sigeti, Phys. Rev. E, {\bf 52}, 2443 (1995).
\bibitem{b1} A. Bershadskii, EPL, {\bf 88}, 60004 (2009).
\bibitem{mm} J.E. Maggs and G.J. Morales, Phys. Rev. Lett., {\bf 107}, 185003 (2011); Phys. Rev.
 E {\bf 86}, 015401(R) (2012).
\bibitem{b2} A. Bershadskii, arXiv:1803.10139 (2018).
\bibitem{lun} T.S. Lundgren, “Linearly forced isotropic turbulence,” in Annual Research Briefs (Center for Turbulence Research, Stanford), 461 (2003)
\bibitem{rm} C. Rosales and C. Meneveau, Phys. Fluids, {\bf 17}, 095106 (2005).
\bibitem{bdy} D. Bogucki, J.A. Domaradzki and P.K. Yeung, J. Fluid Mech. {\bf 343}, 111 (1997).
\bibitem{wcb} L.P. Wang, S. Chen and J.G. Brasseur, J. Fluid Mech. {\bf 400}, 163 (1999).
\bibitem{dy} D.A. Donzis and P.K.Yeung, Physica D, {\bf 239}, 1278 (2010).
\bibitem{dsy} D.A. Donzis  K.R. Sreenivasan,  P.K. Yeung, Flow Turbulence Combust., {\bf 85}, 549 (2010).
\bibitem{wg} T. Watanabe and T. Gotoh, New J. Phys. {\bf 6}, 40 (2004).
\bibitem{yds} P.K. Yeung, D.A. Donzis and K.R. Sreenivasan, Phys. Fluids, {\bf 17}, 081703 (2005).
\bibitem{s} K.R. Sreenivasan, Phys. Fluids {\bf 27}, 1048 (1984).
\bibitem{sb} K.R. Sreenivasan and A. Bershadskii, J. Stat. Phys., {\bf 125}, 1145 (2006).
\bibitem{b3} A. Bershadskii, arXiv:1512.08837 (2015).
\bibitem{jon}D. C. Johnston, Phys. Rev. B {\bf 74}, 184430 (2006).
\bibitem{d} P. A. Davidson P.A. Turbulence in rotating, stratified and electrically conducting fluids. (Cambridge University Press, 2013).
\bibitem{saf} P. G. Saffman, J. Fluid. Mech. {\bf 27}, 551 (1967).
 \bibitem{js} J.V. Jos\'{e} and E.J. Saletan, Classical Dynamics: A Contemporary Approach (Cambridge
University Press, Cambridge 1998).
\bibitem{cvb} P.L. Carroll, S. Verma and G. Blanquart, Phys. Fluids, 25, 095102 (2013).
\bibitem{my} A. S. Monin and A. M. Yaglom, Statistical Fluid Mechanics, Vol. II: Mechanics of Turbulence (Dover Pub. NY, 2007).
\bibitem{ep} V. Eswaran and S.B. Pope, Comput. Fluids {\bf 16}, 257 (1988).
\bibitem{ss} K.R. Sreenivasan and J. Schumacher, Phil. Trans. R. Soc. A, ({\bf 368}, 1561 (2010).
\bibitem{krk} S.M. de Bruyn Kops, J.J. Riley and G. Kos\'{a}ly, Phys. Fluids, {\bf 13}, 1450 (2001).



\end{thebibliography}
\end{document}